\documentclass[aps,twocolumn,showpacs]{revtex4}
\usepackage{graphics}
\begin{document}
\title{Some comments on the divergence of
perturbation series in Quantum Electrodynamics}
\author{Mofazzal Azam}
\address{Theoretical Physics Division, Central Complex,
         Bhabha Atomic Research Centre\\
         Trombay,Mumbai-400085,India}
%\maketitle
%\vskip .8 in
\begin{abstract}
It has been argued by Dyson that the perturbation theory
in coupling constant in QED can not be convergent.
We find that similar albeit slightly different
arguments lead to the divergence of the
series of $~~1/N_f~~$ expansion in QED.\\
%PACS numbers: 11.15.Pg , 11.15.Bt, 12.20.Ds, 11.10.Gh
\end{abstract}
\maketitle
\vskip .2 in
%PACS numbers: 11.15.Pg , 11.15.Bt, 12.20.Ds, 11.10.Gh
\vskip .2 in
%\newpage
It is more than half a century since Dyson argued that the perturbation
theory in the coupling constant in Quantum Electrodynamics (QED) is
divergent (\cite{dyson}, see also \cite{ark,dunne}). In this paper, first
of all, we recall some of the central points in Dyson's arguments.
We, then, consider QED with large number of flavours (i.e.,large number
of species ) of fermions. Large flavour limit is interesting in its' own
right. It has been used to argue for the existence of Landau singularity
in the leading order in $~1/N_f~$  in QED \cite{azam,land}. It has also
been very successfully used in condensed matter physics to proof many
seminal results in Landau Fermi liquid theory, again, in the leading
order in $~1/N_f~$ \cite{fel,shan,fro}. However, these issues will not
be discussed in this paper. We will only be concerned with the nature of
the series thus obtained. In particular, we will argue that the series
obtained by $~1/N_f~$ expasion in QED, where $~N_f~$ is the number of
flavours of fermions, $~N_f~$ tends to infinity with the product
$e^2 N_f$ held fixed, $e$ being the coupling constant, is divergent
using arguments which are similar but slightly different from those of
Dyson's. Our arguments are based on asymptotic freedom in QFT which
is a new development in the field since the work of Dyson.\\
\par Dyson's arguments for the divergence of perturbation theory in QED
is elegant in its' simplicity. Since our paper is centered around
these arguments, we quote the following paragraphs from his paper :\\
"....~let
\begin{eqnarray}
F(e^2)=a_0+a_1 e^2 +a_2 e^4+...              \nonumber
\end{eqnarray}
be a physical quantity which is calculated as a formal power series in
$e^2$ by integrating the equations of motion of the theory
over a finite or infinite time.Suppose, if possible, that the series...
converges for some positive value of $e^2$; this implies that $F(e^2)$
is an analytic function of $e$ at $e=0$.Then for sufficiently small value
of $e$, $F(-e^2)$ will also be a well-behaved analytic function with
a convergent power series expansion.
\par But for $F(-e^2)$ we can also make a physical interpretation.
~...~ In the fictitious world, like charges attract each other.The potential
between static charges, in the classical limit of large distances
and large number of elementary charges, will be just the Coulomb potential
with the sign reversed.But it is clear that in the fictitious world
the vacuum state as ordinarily defined is not the state of lowest energy.
By creating a large number $N$ of electron-positron pairs, bringing the
electrons in one region of space and the positrons in
another separate region, it is easy to construct a pathological state
in which the negative potential energy of the Coulomb forces is much
greater than the total rest energy and the kinetic energy of the particles.
~......~~. Suppose that in the fictitious world the state of the
system is known at a certain time to be an ordinary physical state with
only a few particles present.There is a high  potential barrier separating
the physical state from the pathological state of equal energy; to
overcome the barrier it is necessary to supply the rest energy for
creation of many particles. Nevertheless, because of the quantum-mechanical
tunnel effect, there will always be a finite probability that in any
finite time-interval the system will find itself in a pathological state.
Thus every physical state is unstable against the spontaneous
creation of many particles.Further, a system once in a pathological state
will not remain steady; there will be rapid creation of more and more
particles, an explosive disintegration of the vacuum by spontaneous
polarization.In these circumstances it is impossible that the
integratation of the equation of motion of the theory over any finite
or infinite time interval, starting from
a given state of the fictitious world, should lead to well-defined
analytic functions.Therefore $F(-e^2)$ can not be analytic and the series
~...~ can not be convergent."
\par The central idea in Dyson's arguments for the divergence
of perturbation theory in coupling constant, as is evidient
from the lengthy quotation above, is that the convergence of the
perturbation theory in coupling constant would lead to the
existence of pathological states to which the normal states of
QED would decay.These pathological states correspond to
states of a  quantum field theory whose vacuum state is unstable.
We consider the $~~1/N_f~~$ expansion series in QED
and argue that its' convergence also leads to the existence
of pathological states to which normal states of large flavour
QED would decay. These pathological states are not the
ones considered by Dyson. These states corresponds to states of an
asymptotically free theory of commuting fermions.\\

The Langrangian of QED with number of flavours, $N_{f}$, is
given by,
\begin{equation}
{\cal L} =  \sum_{j=1}^{N_f} \bar{{\psi}}^j \Big(i\gamma^{\mu} \partial_{\mu}
+ m - e \gamma^{\mu} A_{\mu}\Big) \psi^{j} + \frac{1}{4} F_{{\mu}{\nu}}^2
\end{equation}
where $ \psi^{j}$  and $\bar{{\psi}}^j$
are the   Dirac field and its' conjugate, $j$ is the flavour index,
and $ A_{\mu}$  and
$F_{{\mu}{\nu}}$ are the electromagnetic
potential  and the field strength respectively. We will, ultimately,
be considering cases with both the positive and negative sign of $N_f$, and
therefore, we introduce the notation $|N_f|=sign(N_f)\times N_f$~ for latter
convinience.
The ~$\frac{1}{N_f}~$ expansion
is introduced by assuming that, in the limit
~$|N_{f}|{\rightarrow}{\infty}$, ~$e^{2}|N_{f}|~
=constant= ~\alpha^{2} ~(say)~$ .
Alternatively, instead of the Lagrangian given by Eq.(1), we may
consider the following Lagrangian,
\begin{equation}
{\cal L} =  \sum_{j=1}^{N_f} \bar{{\psi}}^j \Big(i\gamma^{\mu} \partial_{\mu}
+ m - \frac{e}{\sqrt |N_f|}~ \gamma^{\mu} A_{\mu}\Big) \psi^{i} +
\frac{1}{4} F_{{\mu}{\nu}}^2
\end{equation}
With this form of the Langrangian, it is easy to set up Feynman diagram
technique.To each photon and fermion line corresponds their usual propagator.
Each vertex contributes a factor of $\frac{e}{\sqrt{|N_f|}}~$, each fermion
loop contributes a factor of $~(-1)~$ for anticommuting
fermions and a factor of
$~N_f~$ because of summation over fermion flavours.Using these rules, it is
easy to set up $~1/N_f~$ expansion series for any physical observable.
Just as in the case of perturbation theory in the coupling constant, the
expansion  in $~\frac{1}{N_f}~$ allows us to express an observable $~F~$
in the form,
\begin{eqnarray}
F(\frac{1}{N_f})=Q_{0}+\frac{1}{N_f}~Q_{1}+\frac{1}{N_{f}^{2}}~Q_{2}+... ...
\end{eqnarray}
$Q_{0}~,~Q_{1}~,~Q_{2},~... ...~~$ are some  functions of the coupling contant.
Now suppose
that the series converges for some small value of $\frac{1}{N_f}~$ ( large
value of $N_f~$ ), then the observable function
$F(\frac{1}{N_f})$ is analytic for
$\frac{1}{N_f}=0~$ ( $N_f=\infty~$ ).Therefore, we can consider a small
negative value of $~\frac{1}{N_f}~$ ( large negative value of $N_f~$ ) for
which the function is analytic and convergent.In other words, the function
$~F(\frac{1}{N_f})~$ can be analytically continued to small negative value
of $~1/N_f~$ and the series thus obtained will be convergent.\\

\par What is the meaning of negative $N_f~$? What does the convergence
of the analytically continued series above imply?
Before we address these questions, we must mention that,
fermions with negative number of flavours have been considered
earlier. In the context of lattice QCD, fermions with finite
number of negative flavours, called bermions have been
considered before(see \cite{divi,rolf} and references there in).
In reference \cite{azam1}, ( an unpublished preprint of the present
author ), it was considered as an exotic
limit of QED. We try to find the meaning of negative flavour
in the spirit of reference \cite{azam1}.
We, first, calculate the effective coupling constant
for positive as well as negative $~N_f~$ using the
formal $~1/N_f~$ expansion series for the two point Green's function.
The series is assumed to be convergent, and therefore, for sufficiently
large $~N_f~$, one can restrict to the leading order term.The leading
order term is given by the one-loop diagrams which can
easily be eavaluated to obtain the polarization from which one can
read off the effective coupling constant. It is given by,
\begin{equation}
e_{eff}^{2}(\Lambda^2) = {e^2  \over 1- \frac{e^2 N_f}{3\pi |N_f|}
ln \frac{\Lambda^2}{m^2}}
\end{equation}
If $N_f$ is negative,
\begin{equation}
e_{eff}^{2}(\Lambda^2) = {e^2  \over 1+ \frac{e^2}{3\pi}
ln \frac{\Lambda^2}{m^2}}
\end{equation}
From the equation above, we find that in the limit
$\Lambda\rightarrow\infty~$, $e_{eff}^{2}\rightarrow 0$, when
$N_f$ is negative. Therefore, the formal theory that we obtain from the
analytical continuation of $1/N_f~$ ( ~for large $N_f$~) to the small negative
value of $~1/N_f~$, is ( at least formally ) asymptotically free.This
seem to suggest that the physical meaning of the negative sign of
$N_f$ could possibly be traced in the free theory without the interaction
term.
\par We will argue that the  choice
of negative $N_f$ for anticommuting fermions amounts to considering
commuting fermions with positive $N_f$ .
Let us consider the Langrangian
given by Eq.(2) in four dimensional Euclidean space.
The partition function is given by the following functional
integral,
\begin{eqnarray}
{\cal Z}_{ac}=\int DA(x) D\bar{{\psi}}(x) D\psi(x) exp(-\int d^{4}x{\cal L})
\end{eqnarray}
We carry out funtional integration with respect to the anticommuting
fermion fields (grassman variables) and obtain,
\begin{eqnarray}
{\cal Z}_{ac}=&&\int DA(x)~ det^{N_f}(i\gamma^{\mu} \partial_{\mu}
+ m - e \gamma^{\mu} A_{\mu}) \nonumber\\
&&exp(- \frac{1}{4} \int d^{4}x  F_{{\mu}{\nu}}^2)
\end{eqnarray}
However, if one considers fermion fields to be commuting variables, then
it turns out to be functional integration over complex fields, and
we obtain,
\begin{eqnarray}
{\cal Z}_{c}=&&\int DA(x)~ det^{-N_f}(i\gamma^{\mu} \partial_{\mu}
+ m - e \gamma^{\mu} A_{\mu}) \nonumber\\
&&exp(- \frac{1}{4} \int d^{4}x  F_{{\mu}{\nu}}^2)
\end{eqnarray}
This expression could be obtained from the previous expression,
simply by assuming that  $N_f$ is  negative.Therefore, anticommuting
fermions  with negative $N_f$ has the same partition function as
the commuting fermions with positive $N_f$. Since, physically
interesting observables can be calculated from the partition function,
our claim is that the negative flavour anticommuting fermions are
equivalent to the positive flavour commuting fermions.
\par This can also be argued using formal perturbation theory.
Consider the two point Green's functions for the photons using Lagrangian
given by Eq.(2) . First, we consider just one loop diagram and show how the
contribution due to flavours appears in the calculations.
There are two vertices and a fermion loop,
each vertex contributes a factor of $~\frac{e}{\sqrt{|N_f|}}~$,
the fermion loop contributes a multiplicative factor of $~(-1)~$
because the fermions anticommute
and a multiplicative factor of  $N_f$ because of summation
over flavours of the internal fermion lines.Now, if $N_f$ happens to be
negative, the factor $~(-1)~$ and  the factor $N_f~$, combines to give the
factor $~|N_f|~$.This is also the contribution if  the fermions commute and
the flavour is positive ( the factor $~ (-1)~$ is
absent for commuting fermions).
The same procedure applies for the multiloop  diagrams.
Calculation of any observable in QED, essentially amounts
to calculating a set of Feynman diagrams.
The information regarding the anticommuting nature
of the fermions in the calculation of the Feynman diagrams
enters through the multiplicative factor of $~(-1~)$ for each fermion
loop that appear in the diagram. Each such loop, as explained above
also contributes a factor of $N_f$. Therefore, when $~N_f~$ is taken
to be negative, the over all multiplicative factor becomes
$~|N_f|~$. As explained above, we would obtain the same multiplicative
factor if we treat the fermions as commuting fields and consider
$~N_f~$ to be positive . This shows that the choice
of negative $~N_f~$ for anticommuting fermions ammounts to considering
commuting fermions with positive $~N_f~$.
We argued above, on formal
grounds, that QED with anticommuting fermions and negative value
of $~N_f~$ is asymptotically free. Therefore, formally speaking,
QED with commuting fermions and positive $~N_f~$ is
asymptotically free \cite{azam1}.

\par It is well known that the free field theory of commuting
fermions does not have a stable vacuum or ground state \cite{gel,nai,wein}.
From the arguments above, it then follows that the interacting theory
also can not have a stable vacuum state.All states in this theory
are pathological. All these results follow from the single assumption
that the $1/N_f$-expansion series in QED is convergent.Therefore,
the convergence of the series in QED with anticommuting
fermions would leads to the
decay of normal states to the pathological states of QED with
commuting fermions via the process of quantum mechanical tunnelling.
Therefore, for QED to be meaningful , the  series in $\frac{1}{N_f}$ ~
expansion must diverge.\\

\vskip .1 in
Acknowledgement : I would like to thank Dr. Satchidanada Naik,
Harishchandra Research Institute, Allahabad, India, without whose
comments, criticism and suggestion this work could not have been completed.
I would also like to thank the annonymous referee for pointing out the
references \cite{divi} and \cite{rolf}.
%\begin{thebibliography}{99}

\end{document}